\documentclass[a4paper,12pt]{article}
\usepackage{eurosym}
\usepackage[utf8]{inputenc}
\usepackage{cancel}
\usepackage{ulem}
\usepackage{amsfonts}
\usepackage{amssymb}
\usepackage{graphicx}
\usepackage{amsmath}
\usepackage{enumerate}
\usepackage{mathtools}
\usepackage{subfig}
\usepackage{color}
\usepackage{tikz}
\usepackage{float}
\usepackage{here}
\usepackage{cite}
\usepackage{mathrsfs}
\usepackage{float,epsfig}
\usepackage{dcolumn}
\usepackage[toc,page]{appendix}
\usepackage{graphicx}
\usepackage{bm}
\usepackage{amsmath,amssymb,amsthm}
\usepackage[colorlinks=true,linkcolor=blue,citecolor=red]{hyperref}
\usepackage{multirow}

\usetikzlibrary{calc}
\newcommand{\be}{\begin{equation}}
\newcommand{\ee}{\end{equation}}
\newcommand{\bea}{\setlength\arraycolsep{2pt} \begin{eqnarray}}
\newcommand{\eea}{\end{eqnarray}}

\def\0{{\sst{(0)}}}
\def\1{{\sst{(1)}}}
\def\2{{\sst{(2)}}}
\def\3{{\sst{(3)}}}
\def\4{{\sst{(4)}}}
\def\5{{\sst{(5)}}}
\def\6{{\sst{(6)}}}
\def\7{{\sst{(7)}}}
\def\8{{\sst{(8)}}}
\def\sst#1{{\scriptscriptstyle #1}}

\makeatletter \@addtoreset{equation}{section}

\definecolor{lime}{HTML}{A6CE39}

\tikzset{>=latex}

\begin{document}

\title{\Large \textbf{  On     Computational CUDA Studies of   Black Hole Shadows }}

\author{\small
S.~E.~Baddis$^{1}$, 
A.~Belhaj$^{1}$, 
H.~Belmahi$^{2}$, 
S.~E.~Ennadifi$^{3}$, 
M.~Jemri$^{1}$\thanks{Authors are listed in alphabetical order. } \thanks{Corresponding author:   maryem.jemri@um5r.ac.ma. } \\[2mm]
{\small $^{1}$ESMaR, Faculty of Science, Mohammed V University in Rabat, Rabat, Morocco} \\
{\small $^{2}$National School of Applied Sciences (ENSA), Chouaib Doukkali University, El Jadida, Morocco} \\
{\small $^{3}$LHEP-MS, Faculty of Science, Mohammed V University in Rabat, Rabat, Morocco}
}

\maketitle

\begin{abstract}
Combining  high-performance CUDA numerical codes with  the  Hamilton--Jacobi formalism, we investigate the shadows  properties of rotating charged Euler--Heisenberg black holes in the presence of global monopoles.   Then,  we discuss  the   associated energy emission rate  by varying the involved   black hole parameters.  As a result,  we  show that both the shadow structure and the energy emission rate depend on the global monopole parameter, the electric charge, and the rotation parameter. However, we observe  that  the Euler--Heisenberg nonlinear parameter   does not significantly affect either the shadow or the energy emission rate. In order to reconcile the present theoretical predictions with
the shadow observations reported by the   Event Horizon Telescope   collaboration, we employ a CUDA-based computational approach
to establish strict bounds on the GM parameter, the electric charge,
and the rotation parameter. {\noindent} 

\end{abstract}

%

\newpage

\section{Introduction}
In black hole physics, a  considerable interest has been devoted to the study of 
 thermodynamic properties. These include the  entropy,  the temperature,  the phase
transitions, and the critical phenomena \cite{f2,f3}. These studies reveal deep
connections between gravity, quantum mechanics, and statistical physics.
Consequently, black holes remain an important subject in the search for a
unified description of fundamental interactions.

Beyond thermodynamic aspects, the optical properties of black holes have
also attracted significant attention \cite{f5,f6,f7}.  Precisely,  the
propagation of light in strong gravitational fields has been widely studied
using analytical and numerical techniques. These analyses have  led to observable
predictions, such as gravitational lensing and the formation of black hole
shadows. A major breakthrough has been  achieved with the observations of the
Event Horizon Telescope (EHT) \cite{f10,f11,f12,f13,f14}. This collaboration has 
produced the first image of the shadow of a supermassive black hole. These
observations strengthen the connection between theoretical models and
astrophysical empirical investigations.

In this regard, charged black holes have attracted remarkable  interest in
recent years. In particular, those described by the Reissner--Nordstr\"om (RN) solution and its extensions have been widely investigated \cite{f31}.
Studies of their shadows show that the presence of  the electric charge can
introduce asymmetries compared to the shadow of a Schwarzschild black hole
with the same mass \cite{f32}. These effects may allow to estimates  the
charge-to-mass ratio, although astrophysical black holes are generally
expected to have a negligible total charge. Shadow properties can also be
modified in other theoretical models. This includes scenarios of frameworks
based on the Euler-Heisenberg theory involving global monopoles (GM). The
latters are considered as relics of early-universe phase transitions, and their
presence in black hole solutions allows to study the  corresponding  interactions 
\cite{f133,f134}. They result from a spontaneous global symmetry breaking
during phase transitions in the early universe. The Goldstone fields
associated with GMs exhibit a slowly decaying energy density behavior,
leading to a linear infrared divergence of the total energy. Such a large
energy distribution can produce significant gravitational effects, modifying
the spacetime geometry. Two classes of such monopoles have been broadly
explored being the ordinary global monopoles (OGM), possessing positive
kinetic energy terms, and the  phantom global monopoles (PGM), which are
characterized by negative kinetic energy terms. In such cases, both the size
and the shape of the shadow can change. These studies are important because
they connect observable shadow features to the internal parameters of black
holes and provide a way to test gravitational theories beyond general
relativity.

Recently, black holes have also been studied in models involving modified
derivatives \cite{cit,f4}. In particular, the Schwarzschild black holes have
been modified by introducing Dunkl derivatives. Within this framework, the
shadows of non-rotating black holes have been investigated, showing that
these modifications can affect their optical behaviors \cite{cit1}. More
recently, strict constraints on the Dunkl deformation parameters have been
derived from the study of the shadows of rotating black holes in the Dunkl
spacetime, using CUDA-based numerical calculations \cite{f34}. This approach
attempts to establish a correlation with the shadow observations reported by
the EHT collaboration. It has been remarked that CUDA has proved to be a powerful platform for
general-purpose parallel computing. It leverages the massive parallelism of
NVIDIA GPUs \cite{43,44}. By distributing computational tasks among
several multiprocessors in a continuous stream,  GPUs enable efficient and
scalable computations. They also provide advanced tools which improve the
performance of the kernel and the overall efficiency of the computations. In
black hole physics, CUDA-accelerated simulations have become particularly
useful. They considerably reduce computation time, improve numerical
stability, and enable more accurate theoretical predictions \cite{s1,s2}.
These techniques are now widely used in studies of black hole shadows. They
also allow for detailed comparisons between theoretical models and
observational data acquired by the EHT.

The aim of this work is to provide a  computational CUDA  study  of   black H-hole shadows the shadow of rotating charged
Euler-Heisenberg black holes with GM using  high-performance
parallel computations.   Combining such numerical methods  with the Hamilton-Jacobi formalism, we analyze the shadow of the rotating
configuration and investigate  the associated   energy emission rate.   In order to conciliate current theoretical predictions with
the shadow observations reported by the EHT  international collaboration, we  explore  a CUDA-based computational approach
to establish strict constraints on the GM parameter $b$, the electric charge $Q$,
and the rotation parameter $a$.

The organization of the paper is as follows. In Section 2, we present the
rotating charged Euler-Heisenberg black holes with the presence of GM.
Section 3 is devoted to the study of the shadow of these rotating charged
black holes using CUDA techniques. In Section 4, we numerically estimate the
energy emission rate to analyze the corresponding Hawking radiation. Section
5 connects the theoretical predictions with observational data from the EHT
collaboration through CUDA-based computations. Finally, the last section
provides some concluding remarks.

\section{ CUDA computations of   shadows of rotating charged Euler--Heisenberg black holes with GMs
}

In this section, we  elaborate  a   CUDA study of   rotating charged Euler-Heisenberg black holes
with GMs by employing the
Newman-Janis algorithm without complexification \cite{21}. This allows the
metric to be expressed as in \cite{22,23}. Based on such an algorithm, we
could investigate  certain  physical properties of the rotating version of the
charged Euler-Heisenberg black holes with GMs using CUDA computations. These GMs, being an example
of topological defects arising from the spontaneous breaking of a global
symmetry $O(3)\ $ to $U(1)$, have a non-ordinary matter effect contribution
leading to a modification of the global structure of spacetime around black
holes. The key physical effect of such a GM contribution is a deficit in the
solid angle of spacetime $\Delta \Omega _{GM}$. To unveil the corresponding
effect, one needs  to introduce the scalar field responsible for the GM. In the
unit system where $G=\hbar =c=1$, the corresponding dynamics could be
described by the total action written as 
\begin{equation}
S\simeq \int d^{4}x\,\sqrt{-g}\left[ \frac{\mathcal{R}}{16\pi }-\frac{1}{4}%
F_{\mu \nu }F^{\mu \nu }+b\left( F_{\mu \nu }F^{\mu \nu }\right) ^{2}+%
\mathcal{L}_{GM}\right]   \label{1}
\end{equation}%
where $R$ is the Ricci scalar and  $g$ represents the metric determinant. The quantity $-%
\frac{1}{4}\left( F_{\mu \nu }\right) ^{2}+b\left( F_{\mu \nu }F^{\mu \nu
}\right) ^{2}$ denotes the Euler--Heisenberg electrodynamics lagrangian, and 
$\mathcal{L}_{GM}$ is the GM lagrangian. The latter can be
described by the scalar lagrangian form 
\begin{equation}
\mathcal{L}_{GM}=\frac{1}{2}\left( \partial _{\mu }\phi ^{a}\right) ^{2}-%
\frac{\lambda }{4}\left( \left( \phi ^{a}\right) ^{2}-\eta ^{2}\right)
^{2}+\mu \mathcal{R}\left( \phi ^{a}\right) ^{2}  \label{2}
\end{equation}%
where $\phi^a$ ($a=1,2,3$)  is the scalar triplet field giving rise to the GM and  $%
\eta $  denotes the corresponding symmetry breaking scale. $\lambda $ denotes the
self-coupling constant, and $\mu $ is a non-minimal coupling constant
between the scalar field and the gravity model. Owing to the fact that the curvature
term $\mu \mathcal{R}\left( \phi ^{a}\right) ^{2}$ slightly shifts the
scalar field vacuum expectation value, we take, for simplicity, the
potential minimum occurring at $\left( \phi ^{a}\right) _{vev}^{2}\equiv \xi
\eta ^{2}$ where $\xi $ is a dimensionless parameter that encodes now the
curvature effect. This vacuum geometry represents a two dimensional real sphere  $\mathbf{S}^{2}$
manifold and the associated symmetry breaking allows topological point
defects, e.g., GMs. Physically, a typical field configuration describing a
GM could be expressed as 
\begin{equation}
\phi ^{a}= \eta h\left( r\right) \frac{x^{a}}{r}, \quad h(r)\mid_{r\to 0} =0,\text{ }h( r)\mid_{r \to +\infty}
=1   \label{3}
\end{equation}%
where one has used $r=\sqrt{x^{a}  x_{a}}$  and $h\left( r\right) $ is a profile
function. Far from the GM core ($r\gg $core size), the above field
configuration becomes 
\begin{equation}
\phi ^{a}\simeq \eta \frac{x^{a}}{r}, \quad    \mid \phi  \mid= \eta  \label{4}
\end{equation}%
where the magnitude of the field is constant. However, its direction changes
with the  position. To reveal the GM influence on black holes, one needs to 
elucidate the behavior of the total energy of the GM. Such an energy, coming
mainly from the kinetic term of Eq.~(\ref{3}), diverges linearly with distance.  Using  the spherical coordinates, indeed,  we have 
\begin{equation}
E_{GM}=\int \frac{1}{2}\left( \partial _{\mu }\phi ^{a}\right)
^{2}d^{3}x\simeq \int_{0}^{R}\left( \frac{\eta }{r}\right) ^{2}\left( 4\pi
r^{2}\right) dr\sim 4\pi \eta ^{2}R  \label{5}
\end{equation}
where $R$ refers to  distance cutoff. It is an energy distribution extension
of the GM  generating  an unusual gravitational field affecting the
surrounding spacetime. Considering this GM energy behavior and
solving the corresponding Einstein equations can  result in a constant shift in
the metric function, giving the deficit solid angle term 
\begin{equation}
\Delta \Omega _{GM}=8\pi \eta ^{2}\xi .  \label{6}
\end{equation}
This spacetime geometry modification generated by GMs  affects the
metric functions of the involved black holes. Concretely, adopting the Boyer-Lindquist coordinate system, we can obtain the metric functions of the
involved black holes. Indeed, we can get the following line element for the black hole metric

{\footnotesize 
\begin{equation}
ds^{2}=\left( \frac{\sigma (r)}{\Sigma (r)}-1\right) dt^{2}-\frac{%
2a^{2}\sigma (r)}{\Sigma (r)}\sin ^{2}\theta \,dt\,d\phi +\left( r^{2}+a^{2}+%
\frac{a^{2}\sigma (r)\sin ^{2}\theta }{\Sigma (r)}\right) \sin ^{2}\theta
\,d\phi ^{2}+\frac{\Sigma (r)}{\Delta (r)}dr^{2}+\Sigma (r)d\theta ^{2},
\label{mr}
\end{equation}
}
where $a$ is the rotating spin parameter. The corresponding metric
functions are given by 
\begin{equation*}
\Sigma (r)=r^{2}+a^{2}\cos ^{2}\theta ,\quad \Delta
(r)=r^{2}f(r)+a^{2},\quad \sigma (r)=r^{2}-r^{2}f(r)
\end{equation*}%
where the GM effect, e.g., the  above deficit solid angle term, appears in the
metric function $f(r)$ as  
\begin{equation}
f(r)=1-\frac{2M}{r}+\frac{Q^{2}}{r^{2}}-\frac{bQ^{4}}{20r^{6}}-8\pi \eta
^{2}\xi .
\end{equation}%
The quantities $M$ and $Q$ denote the total mass and the electric charge
 of the black hole, respectively. The parameter $b$ is the
Euler-Heisenberg parameter \cite{i1}. Though it is positive in the original
Euler-Heisenberg formulation, it is significant in gravitational theories to
also consider negative values in the form of a nonlinear and independent
parameter. At this point, we would like to make a few comments. It has been
remarked that the GM term $8\pi \eta ^{2}\xi $ appears in optical and
thermodynamic quantities of various black hole solutions \cite{24}. Removing
the external parameters required by $b=\eta =\xi =0$, we recover to the
standard charged  black hole solution \cite{25}.

\begin{figure}[h!]
\begin{center}
        \includegraphics[width=4cm,height=4cm]{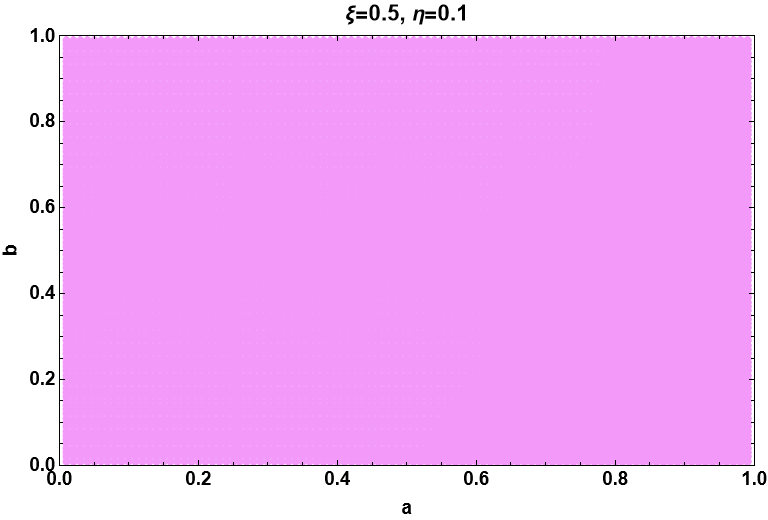} 
        \includegraphics[width=4cm,height=4cm]{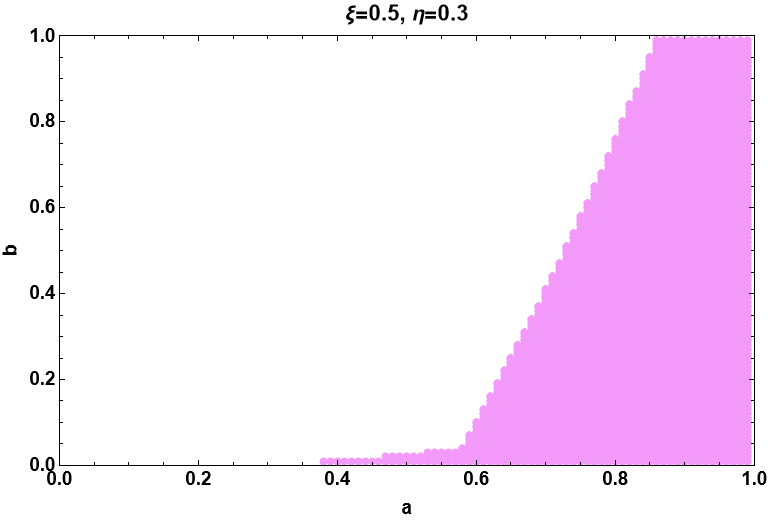}
        \includegraphics[width=4cm,height=4cm]{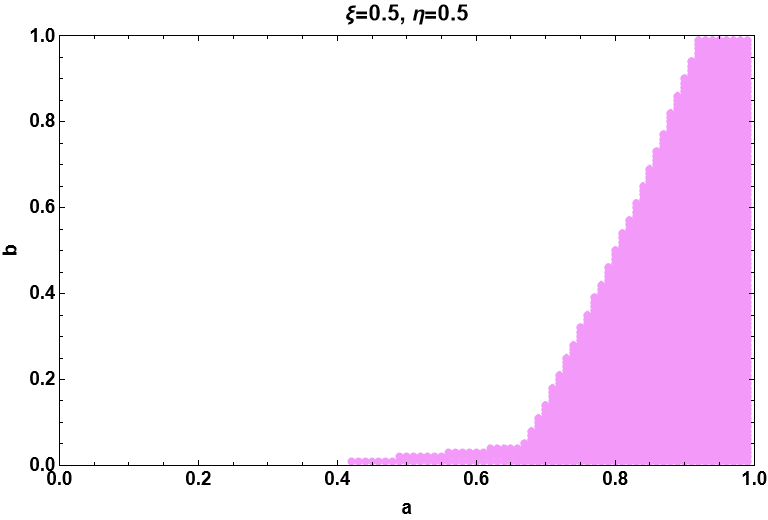}
        \includegraphics[width=4cm,height=4cm]{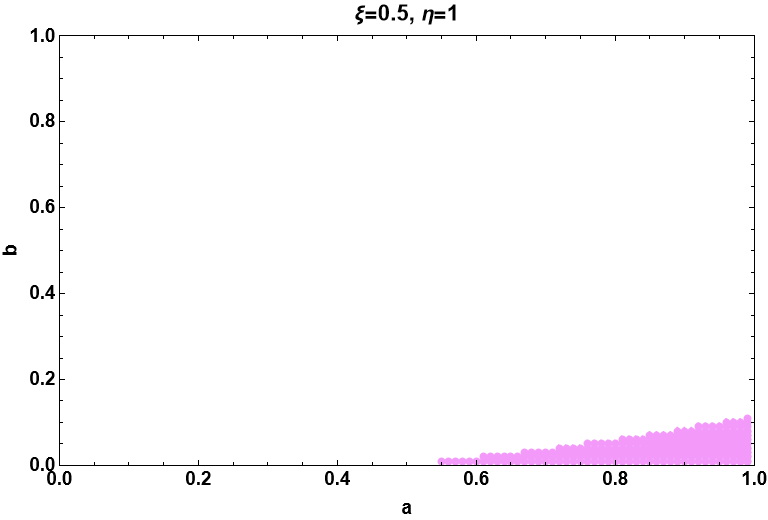}

        \includegraphics[width=4cm,height=4cm]{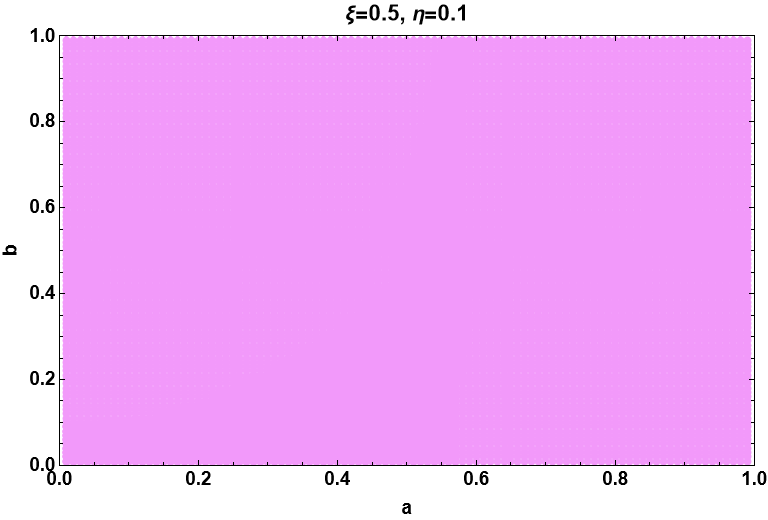} 
        \includegraphics[width=4cm,height=4cm]{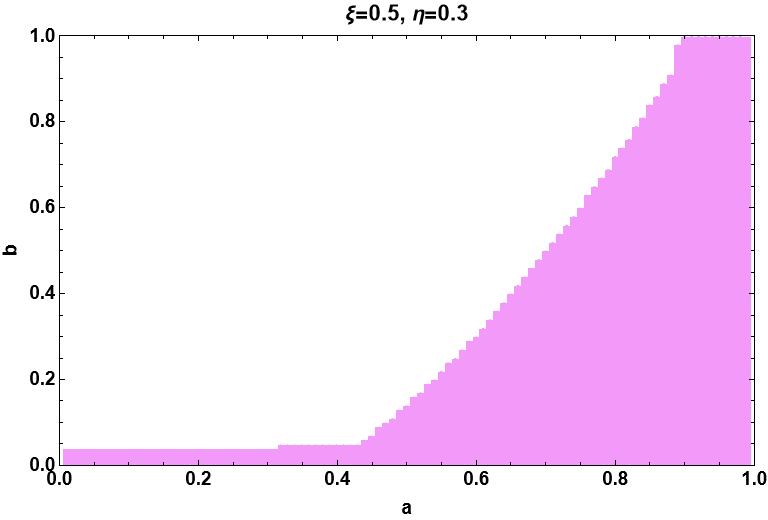}
        \includegraphics[width=4cm,height=4cm]{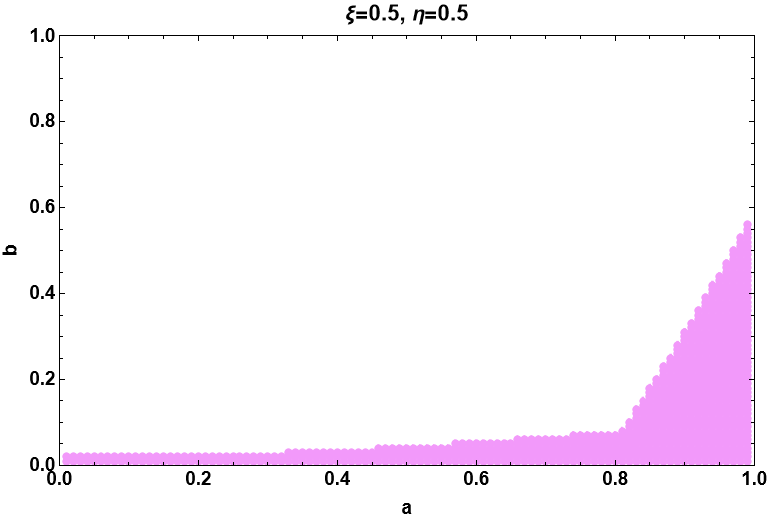}
        \includegraphics[width=4cm,height=4cm]{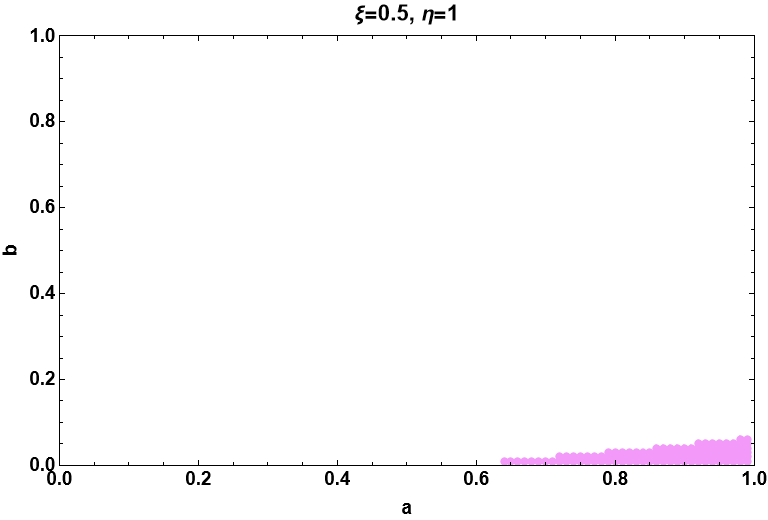}   
\end{center}
\caption{Regions in the $(b, a)$--plane, where the metric admits at least
one real event horizon radius.}
\label{im1}
\end{figure}

Having discussed the metric function, we  analyze its possible horizons using a CUDA-based numerical implementation that enables efficient parallel computation on GPUs. These horizons are determined by  vanishing  the radial component of the metric. To determine the existence of horizons, we implement a numerical algorithm in which the parameters $\xi$, $Q$ and $\eta$ are kept fixed, while $b$ varies from $0$ to $1$ with a step size of $0.1$. Moreover, $a$ ranges with the same increment. For each pair $(a,b)$, the horizon equation is solved numerically in order to show the presence of at least one real solution, corresponding to a physical horizon. This procedure allows one to consistently identify regions of the parameter space where the metric admits such solutions. To illustrate this scenario,  Fig.~(\ref{im1}) displays the regions in the $(a,b)$ parameter space where at least one real horizon exists (at $\xi = 0.5$) for several values of $\eta$ by considering two charge configurations: $Q = 0.4$ (top panel) and $Q = 0.9$ (bottom panel).\\
The figure shows that an increase in the charge value reduces the allowed
region where the black hole horizons exist. Interestingly, for the case $\eta =
0.1$ and $\xi = 0.5$, the charge variation has no effect, and the entire
region remains allowed. On the other hand, increasing $\eta$ substantially shrinks the allowed
region, which becomes very narrow in the case $\eta = 1$. Indeed, the following optical study will be carried out by considering the
region of the moduli parameter space aligned with the allowed domain 
admitting  at least one real solution.

In this section, we examine the shadow of a rotating charged Euler
-Heisenberg black hole with a GM. Precisely, we analyze the effect of each
parameter on the black hole shadow. To perform this evaluation effectively,
we employ a CUDA-based numerical code which enables high-performance
parallel computations. It is recalled that CUDA is a general-purpose
parallel computing platform and programming model that leverages the
parallel computing capabilities of NVIDIA GPUs. In particular, the GPU
architecture allows workloads to be distributed among a large number of
streaming multiprocessors (SMs) in a highly parallel manner. In addition,
modern GPUs provide a variety of powerful methods and efficient tools to
further exploit this architecture. With each new generation, improvements in
CUDA core design strategies lead to increased performance and computational
efficiency. Moreover, CUDA has proven to be a powerful tool in black hole
investigations enabling efficient GPU-accelerated simulations \cite{s1,s2}.
This significantly reduces computation time, improves numerical stability,
and facilitates the exploration of the theoretical predictions in black hole
physics \cite{s3,s4}. To start, it is denoted that the shadow of a black
hole is defined as the apparent boundary, or the critical curve, observed when
light rays asymptotically approach an unstable circular orbit known as the
photon sphere and then return toward the observer. This behavior is encoded
in the null geodesics around black holes. To study the optical properties of
the rotating black holes, it is necessary to establish specific relations using
the Hamilton-Jacobi formalism. In particular,
the separation of variables can be performed through the Carter method \cite%
{26}. For such black hole solutions, the four equations of motion can be
written as follows 
\begin{align}
\Sigma \dot{t}& =\frac{r^{2}+a^{2}}{\Delta }\left[ E\left(
r^{2}+a^{2}\right) -aL\right] +a\left[ L-aE\sin ^{2}\theta \right] \\
(\Sigma \dot{r})^{2}& =\mathcal{R}(r) \\
(\Sigma \dot{\theta})^{2}& =\Theta (\theta ) \\
\Sigma \dot{\phi}& =\left[ L\csc ^{2}\theta -aE\right] +\frac{a}{\Delta }%
\left[ E\left( r^{2}+a^{2}\right) -aL\right] ,
\end{align}%
where $E$ and $L$ are the energy and the angular momentum of the light rays,
respectively. $\mathcal{R}(r)$ and $\Theta (\theta )$ functions are
expressed as follows 
\begin{align}
\mathcal{R}(r)& =\left[ E\left( r^{2}+a^{2}\right) -aL\right] ^{2}-\Delta %
\left[ \mathcal{C}+\left( L-aE\right) ^{2}\right] , \\
\Theta (\theta )& =\mathcal{C}-\left( L\csc \theta -aE\sin \theta \right)
^{2}+\left( L-aE\right) ^{2},
\end{align}%
where $\mathcal{C}$ is the Carter separation parameter. Solving the unstable
circular orbit equations, the two needed impact parameters are obtained as
follows 
\begin{align}
\xi & =\frac{r^{2}\!\left[ \,16a^{2}\Delta (r)+8r\,\Delta (r)\Delta ^{\prime
2}-r^{2}\Delta ^{\prime 2}\right] }{a^{2}\,\Delta ^{\prime 2}}\bigg|%
_{r=r_{0}},  \label{3.5} \\[4pt]
\Xi & =\frac{(r^{2}+a^{2})\,\Delta ^{\prime }(r)-4r\,\Delta (r)}{a\,\Delta
^{\prime }(r)}\bigg|_{r=r_{0}}.  \label{3.6}
\end{align}%
For the rotating charged Euler--Heisenberg black holes with GMs, the
apparent shape of the black hole shadow, as observed at spatial infinity,
can be characterized by the celestial coordinates $(X,Y)$ being 
\begin{align*}
X& =\lim_{r_{\text{ob}}\rightarrow +\infty }\left( -r_{\text{ob}}^{2}\sin
\theta _{\text{ob}}\frac{d\phi }{dr}\right) \\[4pt]
Y& =\lim_{r_{\text{ob}}\rightarrow +\infty }\left( r_{\text{ob}}^{2}\frac{%
d\theta }{dr}\right) ,
\end{align*}%
where $r_{ob}$ is the distance of the observer from the black hole. It is
denoted that $\theta _{\text{ob}}$ indicates the angle of inclination
between the line of the observer and the axis of rotation of the black hole.

To explore how each parameter affects the black hole shadow, we exploit a
numerical method to illustrate the shadow curves choosing a wide range of
values.  Precisely, a  CUDA-based parallel computing program is employed to
speed up the calculations ~\cite{43,44} . This approach allows rapid determination of the
shadow boundaries for different parameter variations. For each illustration,
all parameters are kept fixed except the one of interest. This parameter is
incremented in steps of $0.001$. Indeed, we solve Eqs.~(\ref{3.5}) and  Eqs.~(\ref%
{3.6}). Then, we implement the resulting values in the shadow equation. This
process provides a precise assessment of the influence of each parameter on
the shadow geometric deformation including the size and the shape.

\begin{figure}[!ht]
\centering
\includegraphics[scale=0.55]{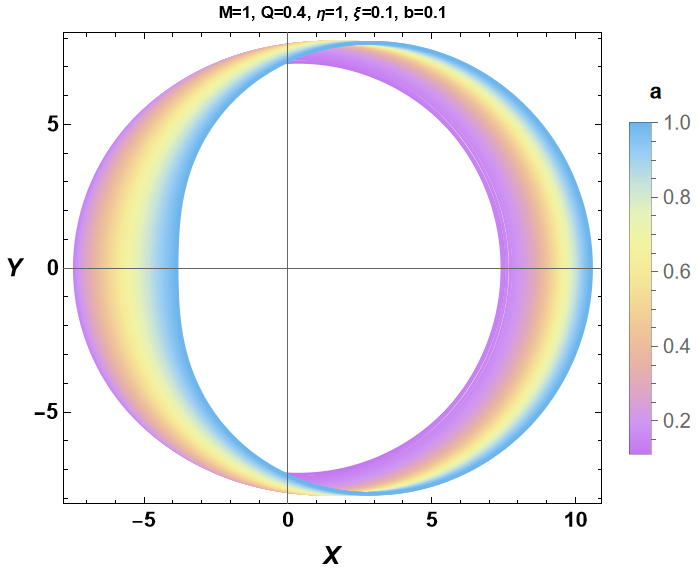} \includegraphics[scale=0.49]{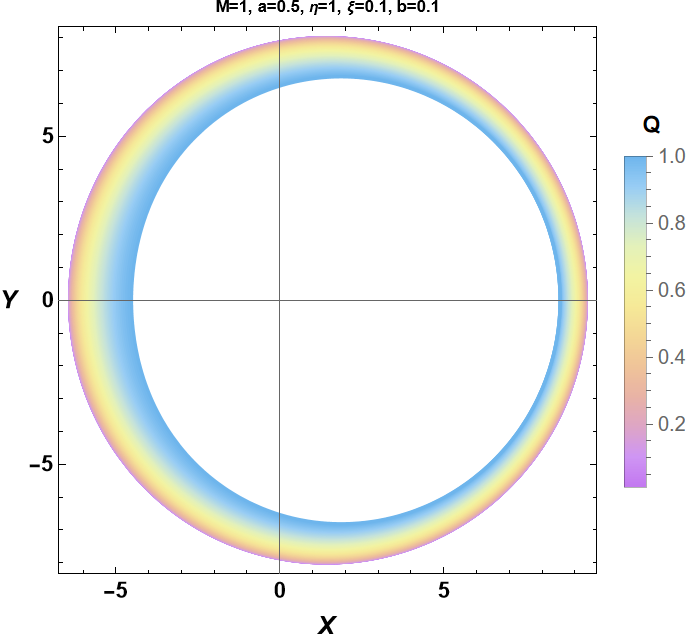}
\caption{\textit{\protect\footnotesize Effect of internal parameter on
shadow behavior.}}
\label{im44}
\end{figure}

In Fig.~(\ref{im44}), we illustrate the shadow appearance by examining the
effects of the charge $Q$, the rotation parameter $a$, where the parameter $b$ 
is fixed at $0.1$. 

As shown in the figure, the rotation parameter behaves similarly to that in
ordinary black holes, slightly reducing the size of the shadow and deforming
its shape into a D-like form. Hence, this parameter retains its role as a
deformation parameter for these black hole solutions. Concerning the charge
effect, increasing the charge reduces the shadow size without significantly
modifying its shape, in agreement with the behavior of standard charged
black holes.   In the present solutions, however, the variation range of the
shadow size extends to values of about $15$, unlike ordinary charged black
hole solutions where the radius does not exceed $7$. This feature arises
from the additional terms in the metric, in which the parameter $b$ is
coupled to the charge $Q$.

For the GM parameters, which are coupled, fixing one while
varying the other reveals their combined effect. Indeed, by taking $\xi =1$
and varying $\eta $, we observe that this contribution increases the size of
the shadow. 
\begin{figure}[th]
\centering
\includegraphics[scale=0.52]{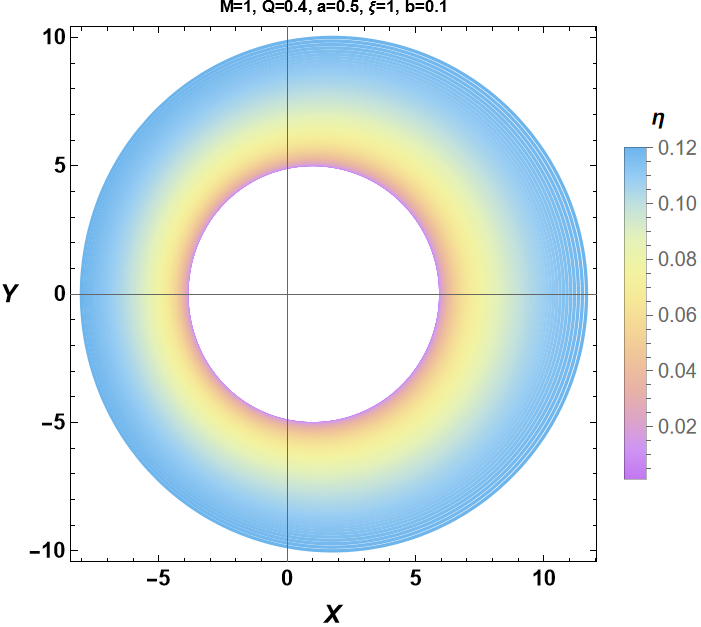} \includegraphics[scale=0.5]{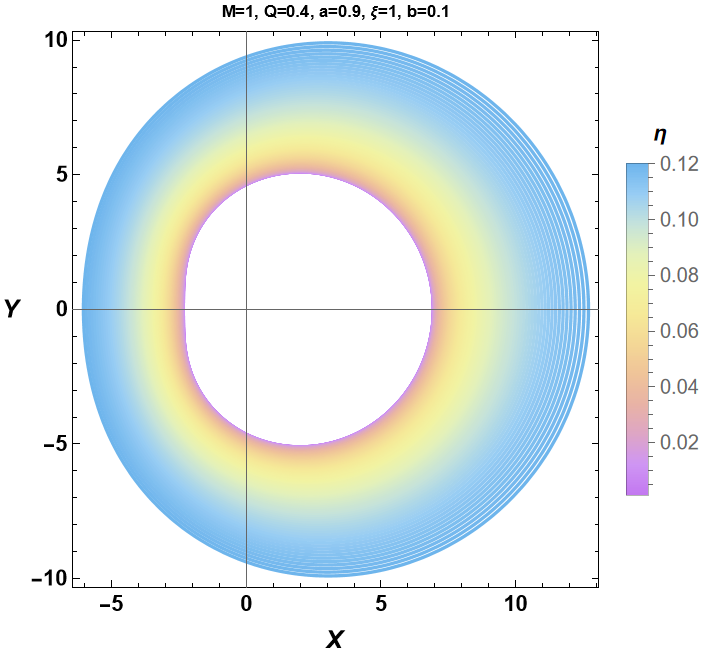}
\caption{\textit{\protect\footnotesize Effect of $\protect\eta $ on shadow
behavior.}}
\label{im4}
\end{figure}
As shown in Fig.~\ref{im4}, interesting behaviors appear, for large values
of the rotation parameter $a=0.9$ and small values of $\eta $, the shadow
exhibits a D-like form. As $\eta $ increases, this D-like form gradually
disappears, illustrating the interplay between the GM and the
black hole rotation. Concretely, this circularization,  (or symmetry
effect) \ appearing as  $\eta $  grows, is due to the fact
that the GM term $8\pi \eta ^{2}\xi $ affects the geometry
globally and isotropically, by  diluting  the relative influence of the spin. 

\begin{figure}[th]
\centering
\includegraphics[scale=0.55]{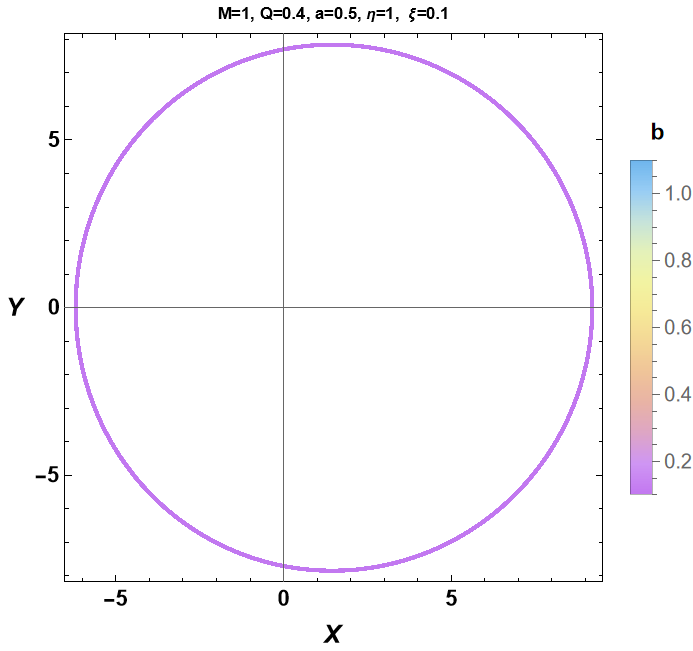} \includegraphics[scale=0.5]{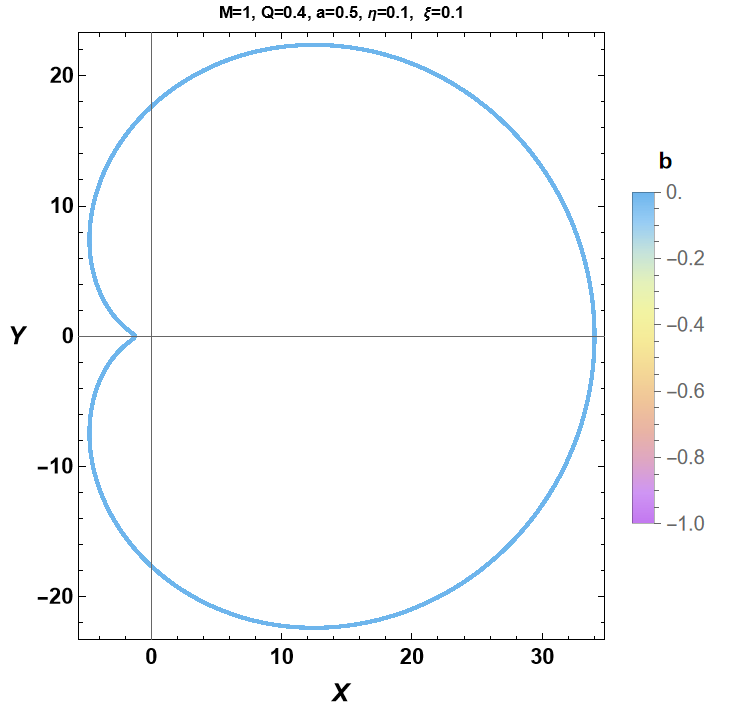}
\caption{\textit{\protect\footnotesize Effect of $b$ parameter on shadow
behavior.}}
\label{im45}
\end{figure}
We  move now to  consider the effect of the $b$ parameter. Fig.~(\ref{im45}) depicts
the black hole shadow for positive and negative values of $b$. The results
indicate that variations in the magnitude of $b$ do not significantly alter
the shadow size.  Interestingly, the 
positive values of $b$ lead to a D-shaped shadow, whereas negative values
produce a cardioid-like configuration.

\section{Energy emission rate using CUDA computations}

In this section, we employ a CUDA-based numerical method to approach the
energy emission rate associated with  the rotating and  the charged Euler-Heisenberg black holes with GM. By leveraging the parallelism of GPUs,
this method provides efficient and high-performance computation of effective
absorption cross sections over a wide range of parameters. To a distant
observer, the cross section for absorption at very high energies
asymptotically trends toward its geometric optical bound, being directly
linked to the size of the shadow of the black hole. In intermediate
operating conditions, the effective absorption cross-section oscillates
around a constant limit value, referred to as $\sigma_{\text{lim}}$. It has
been established that this constant coincides with the geometric
cross-section of the photon sphere, determined by the properties of the null
geodesics \cite{27,28,29}. Since the shadow determines the optical
appearance of the black hole, it can be treated as this limit value, which
can be used to approximate $\sigma_{\text{lim}}$ as follows 
\begin{equation}
\sigma_{\text{lim}} \simeq \pi R_{s}^{2},
\end{equation}
where $R_{s}$ is the shadow radius. Within this framework, the differential
energy emission rate takes the form 
\begin{equation}
\frac{d^{2}E(\omega)}{d\omega \, dt} = \frac{2\pi^{3} R_{s}^{2}}{%
e^{\omega/T_{H}}-1} \, \omega^{3},
\end{equation}
where $T_{H}$ is the Hawking temperature of the black hole and $\omega$ is
the emission frequency. This relation establishes a clear connection between
the thermodynamic properties of the black hole and its optical features.
Indeed, this may provide a useful tool to probe the spacetime parameters
through the observational signatures. Considering the rotating metric, the
Hawking temperature of such black holes is given by 
\begin{equation}
T_{H}=\frac{10 \left(1-8 \pi \,\eta^{2} \xi \right) r^{6}-10 M \,r^{5}+b
\,Q^{4}}{20 \pi r^{5} \left(a^{2}+r^{2}\right)}.
\end{equation}
To evaluate the energy emission rate for different black hole parameter
values, we perform numerical simulations using a CUDA-based program. The
corresponding code computes first the maximal shadow radius from the
obtained shadow data. Then, the horizon radius is determined by solving $%
\Delta(r_h)=0$ and substituted into the Hawking temperature formula. To
assess the impact of each parameter, we vary the parameter of interest in
steps of 0.001 while keeping all others fixed. These results are then used
to generate the energy emission rate plots.

\begin{figure}[!ht]
\begin{center}
\centering
\begin{tabbing}
			\centering
			\hspace{0.5cm}\=\kill
					\includegraphics[scale=0.38]{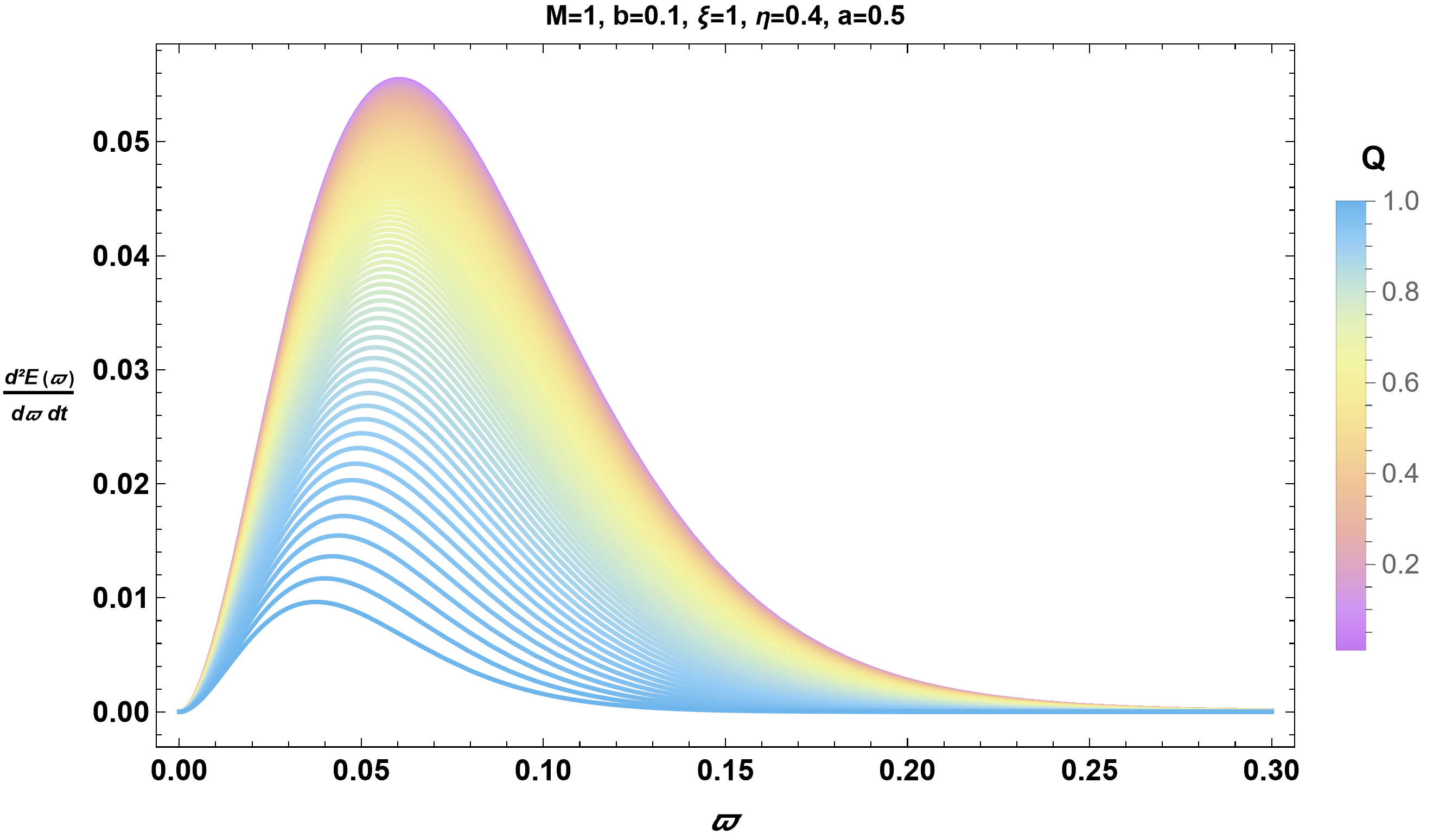}\hspace{0.05cm}	\includegraphics[scale=0.38]{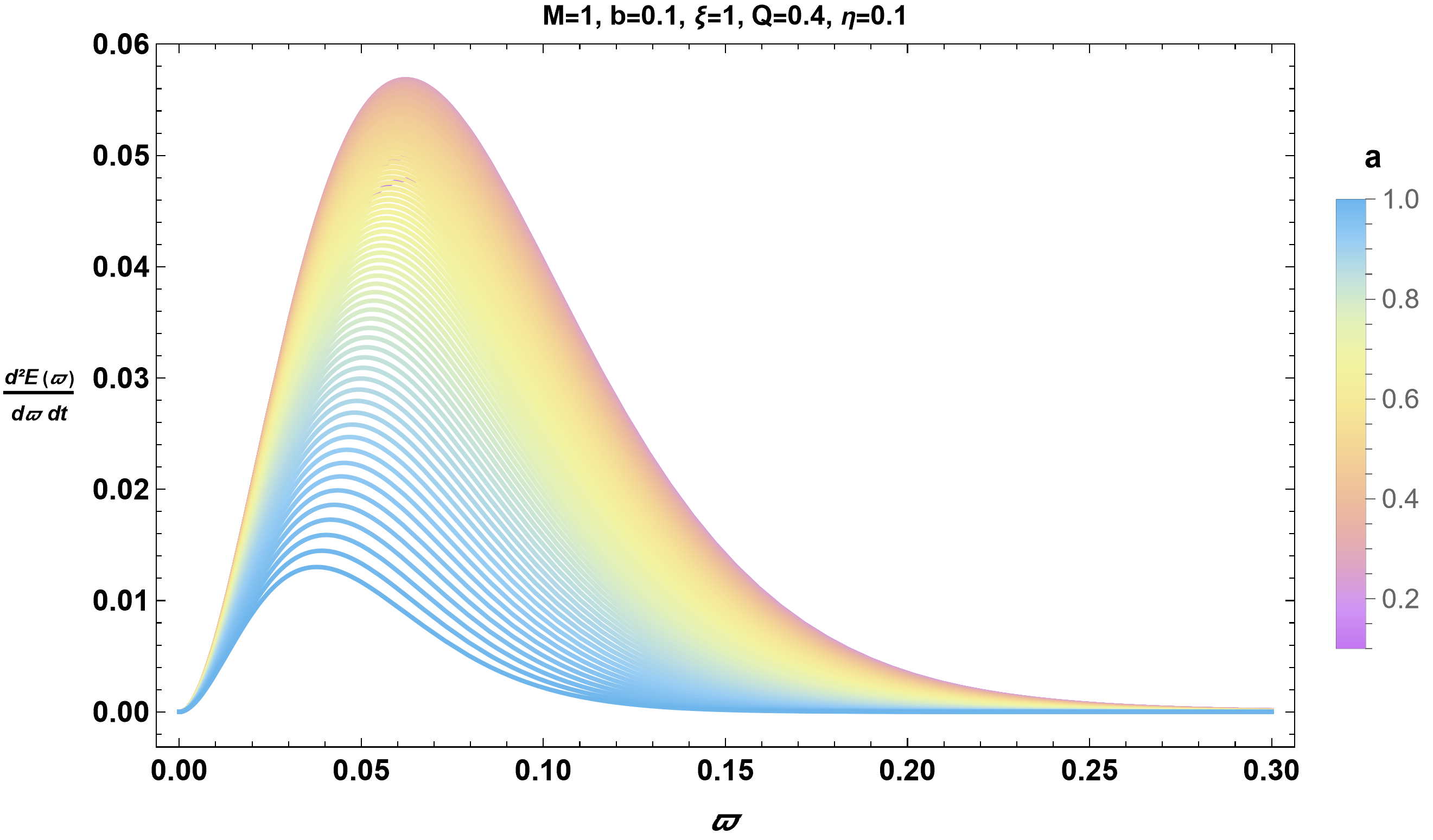}\\ 
	 \end{tabbing}
\end{center}
\caption{ \textit{\protect\footnotesize Variation of the energy emission
rate as a function of the emission frequency for different values of $a$ and 
$Q$. }}
\label{ER}
\end{figure}

In Fig.(\ref{ER}), we illustrate the variation of the energy emission rate
as a function of the emission frequency. This figure represents  the effects of
the electric charge and the rotation parameter on this variation. It is
clear from the figure that increasing the charge leads to a decrease in the
energy emission rate. In contrast, the presence of the rotation parameter
enhances the energy emission rate. Indeed, the charge acts as a suppressing
factor, while the GM behaves as an amplifying contribution,
which is the usual effect of these two parameters.

\begin{figure}[!ht]
\begin{center}
\centering
\begin{tabbing}
			\centering
			\hspace{0.5cm}\=\kill
					
	\includegraphics[scale=0.38]{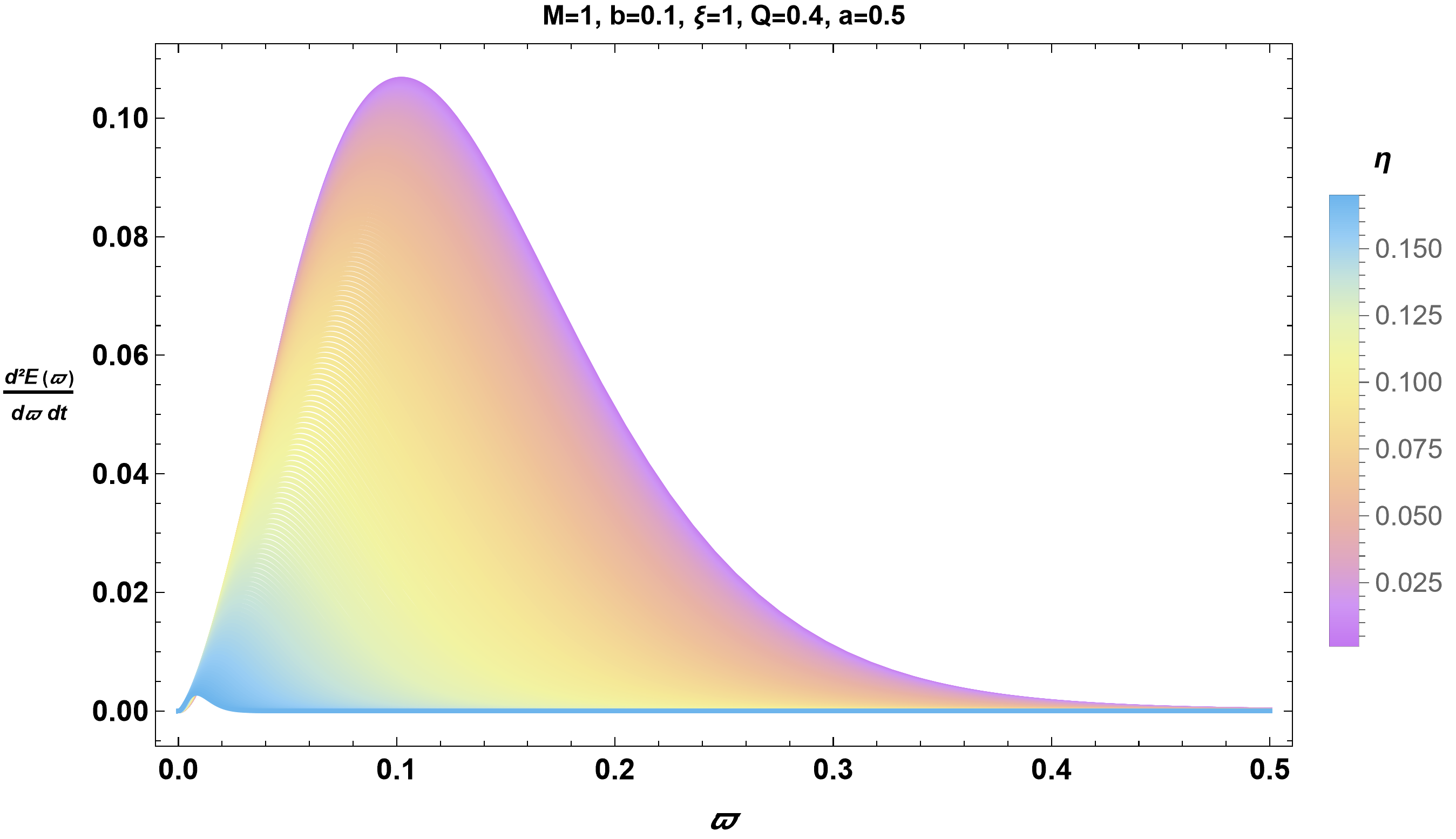} 
	\hspace{0.05cm}		\includegraphics[scale=0.38]{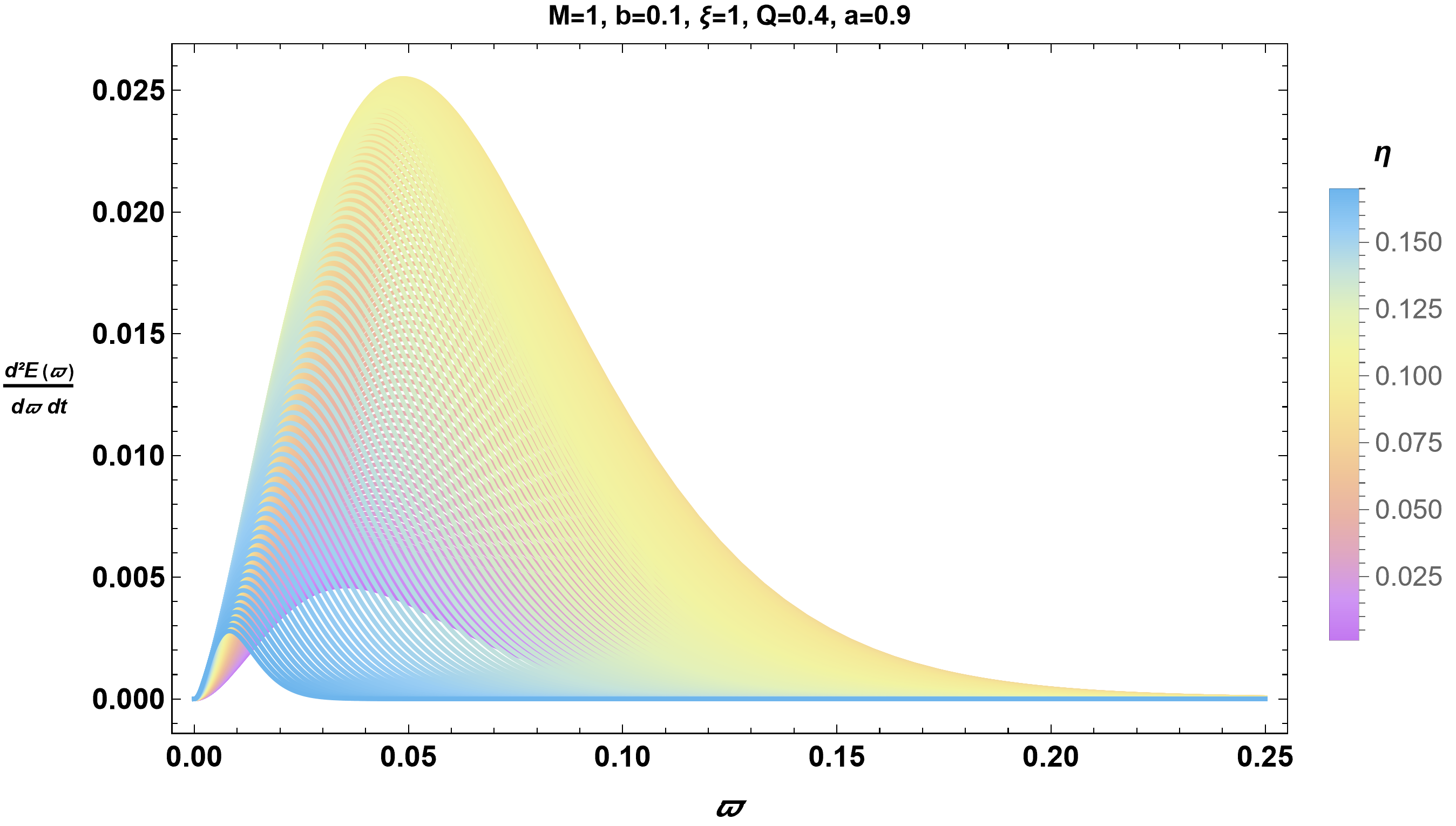}\\  
	 \end{tabbing}
\end{center}
\caption{ \textit{\protect\footnotesize Variation of the energy emission
rate as a function of the emission frequency for different values of $%
\protect\eta$. }}
\label{ER1}
\end{figure}

Fig.~(\ref{ER1}) shows the effect of the GM parameter for both small and
large values of the rotation parameter. For small rotation values,  the GM parameter decreases the energy emission rate.   For large  values, however,  the GM initially increases the energy emission rate up to a critical value, beyond which it begins to act as a
suppressing factor.

\begin{figure}[!ht]
\centering
\includegraphics[scale=0.38]{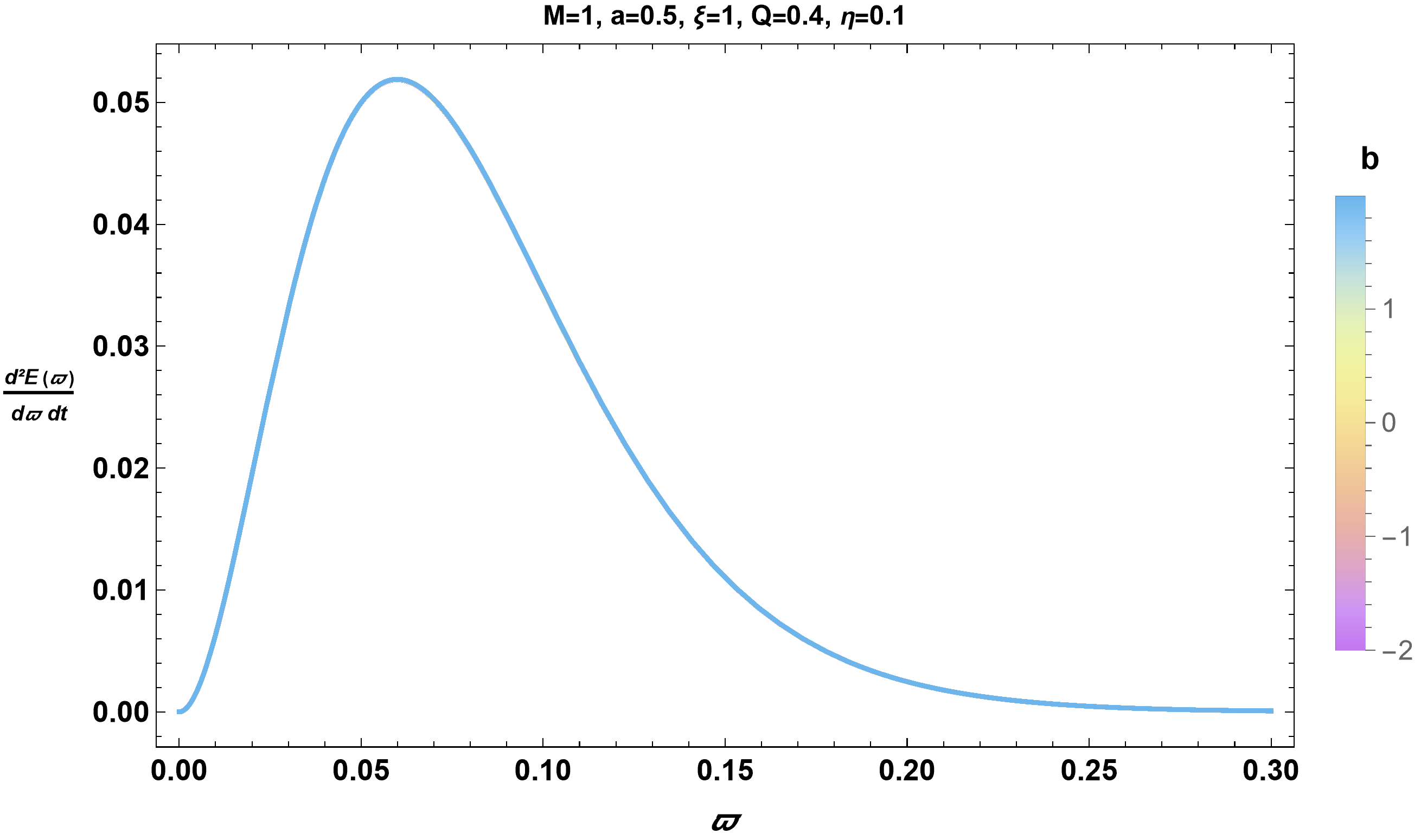}
\caption{ \textit{\protect\footnotesize Variation of the energy emission
rate as a function of the emission frequency for different values of $b$ .}}
\label{Eb}
\end{figure}

We now turn to the effect of the parameter $b$ where the association variation  is  shown in Fig.~(\ref{Eb}).
As discussed in the previous section, this parameter has no significant
impact on the shadow radius. Although the energy emission rate also depends
on the temperature, which varies with the parameter $b$, the figure
indicates that, overall, $b$ has a negligible effect on the energy emission
rate. This behavior holds for both negative and positive values of $b$. 

\section{Constraints on black hole parameters from EHT observations using  CUDA techniques}

In order to establish a bridge between the theoretical predictions and the
observational data, this section provides an analysis of the shadow cast by
rotating Euler-Heisenberg black holes with GMs, in connection
with observational results reported by the EHT collaborations. Concretely,
we exploit the observational data of M87* black hole and Sagittarius A* (Sgr A*) to
constrain the parameters of such black holes~\cite{E1,E2,E3}. The numerical
analysis is performed using a CUDA-based code developed by NVIDIA, which
leverages parallel computing on GPUs to significantly speed up the numerical
calculations required to determine the shadow of the black holes.

Roughly, the constraints can be obtained by using the fractional deviation
from the Schwarzschild black hole shadow diameter expressed by

\begin{equation}
{\ d} = \frac{R_s}{r_{sh}}-1,
\end{equation}
where $R_s $ denotes the shadow radius, $M $ is the mass of the black hole, and $r_{sh}$ denotes the Schwarzschild radius. The dimensionless quantity $R_s/M $
provides a key observable for comparing theoretical models with empirical
measurement results. The 1-$\sigma$ and 2-$\sigma$ confidence intervals
derived from the EHT observations are summarized in Table~\ref{t1}.

\begin{table}[h!]
\centering
\begin{tabular}{|c|c|c|c|}
\hline
\textbf{Black Hole} & \textbf{Deviation ($d$)} & \textbf{1-$\sigma$ Bounds}
& \textbf{2-$\sigma$ Bounds} \\ \hline
M87$^*$ (EHT) & $-0.01^{+0.17}_{-0.17}$ & $4.26 \leq \frac{R_s}{M} \leq 6.03$
& $3.38 \leq \frac{R_s}{M} \leq 6.91$ \\ \hline
Sgr~A$^*$ (EHT$_{\text{VLTI}}$) & $-0.08^{+0.09}_{-0.09}$ & $4.31 \leq \frac{%
R_s}{M} \leq 5.25$ & $3.85 \leq \frac{R_s}{M} \leq 5.72$ \\ \hline
Sgr~A$^*$ (EHT$_{\text{Keck}}$) & $-0.04^{+0.09}_{-0.10}$ & $4.47 \leq \frac{%
R_s}{M} \leq 5.46$ & $3.95 \leq \frac{R_s}{M} \leq 5.92$ \\ \hline
\end{tabular}%
\caption{ \textit{\protect\footnotesize Estimated fractional deviations and
corresponding bounds for M87$^*$ and Sgr~A$^*$ black holes.}}
\label{t1}
\end{table}
In what follows, we provide a numerical algorithm using CUDA-based
computations to determine the parameter pairs $(\eta, Q)$ and $(\eta, a)$
producing  the black hole shadow configurations consistent with the observational
data. Indeed, we fix $\xi = 0.4$, $b = 0.5$, and $M = 1$ throughout the
analysis. First, for each parameter combination, the maximal shadow radius $%
R_{\mathrm{max}}$  has been  computed using the CUDA code, allowing a efficient
evaluation over a dense grid of values. For the $(\eta, Q)$ analysis, $a$ is
set to 0.5 while $\eta$ is varied from 0 to 0.25 and $Q$ from 0 to 1. For
the $(\eta, a)$ analysis, $Q$ is fixed while $\eta$ varies from 0 to 0.25
and $a$ from 0 to 1 with a step size of 0.001. For each combination of
parameters, the computed shadow radius is compared to the observational
bounds reported by the EHT collaboration. This permits  to identify
the allowed regions in the parameter space satisfying the $1\!-\!\sigma$ and 
$2\!-\!\sigma$ confidence intervals.

\begin{figure}[h!]
\centering
\includegraphics[scale=0.45]{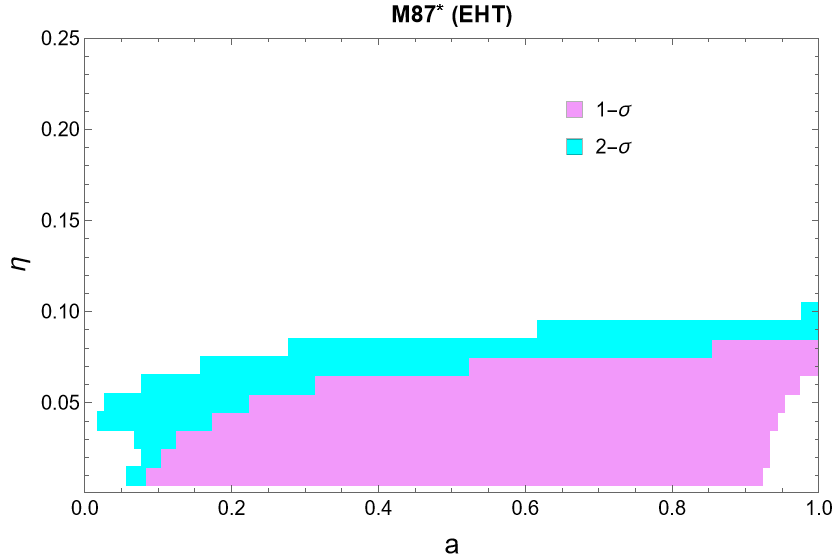}\hspace{2mm} %
\includegraphics[scale=0.45]{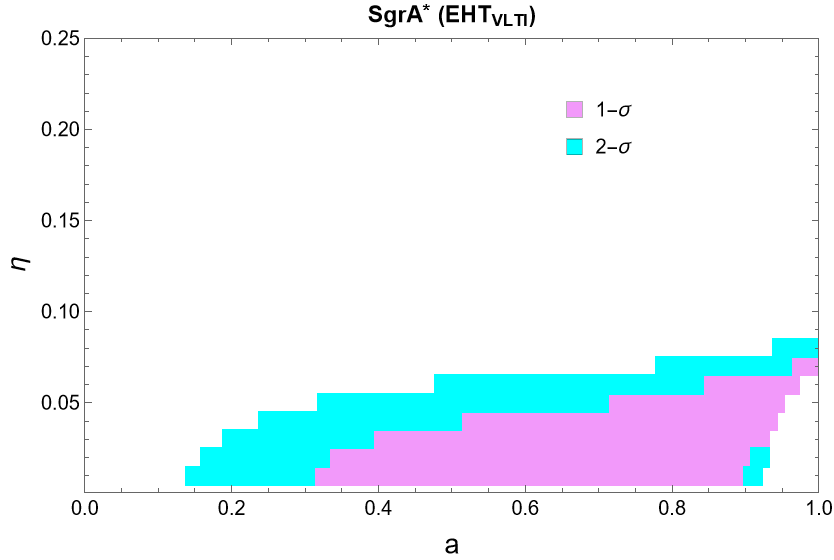} \hspace{2mm} %
\includegraphics[scale=0.45]{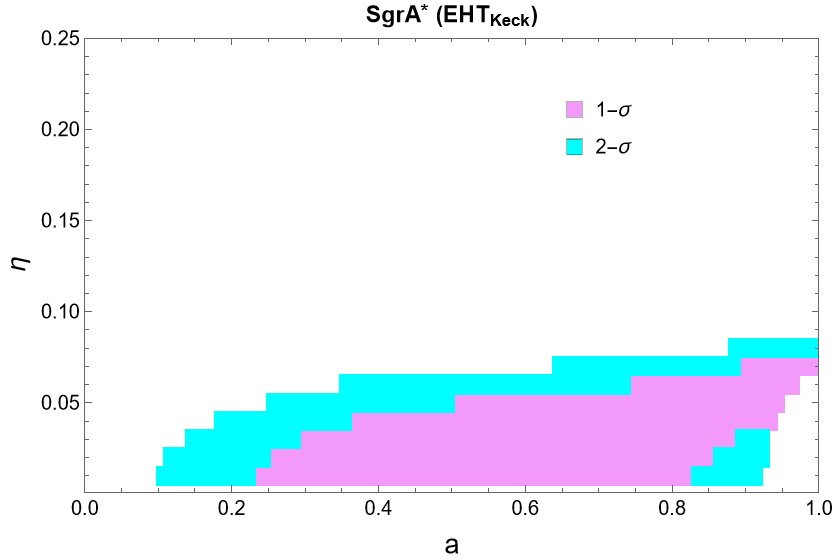}
\caption{ \textit{\protect\footnotesize Constraint Constraint regions in the 
$( \protect\eta,Q)$ plane obtained from CUDA-based simulations, showing
agreement with the EHT observations of M$87^*$ and Sgr~A$^*$ within $1-%
\protect\sigma$ and $2-\protect\sigma$ confidence levels for $Q=0.4$ with $%
M=1$.}}
\label{fig:Moduli}
\end{figure}

As illustrated in Fig.~(\ref{fig:Moduli}), the regions of the 
reduced parameter space $(\eta,a)$ consistent with empirical observations expand for
larger values. This suggests that the black hole spacetime can effectively
reproduce the observed shadow signatures. Due to the correlation between
these parameters, one of them is set while restricting the remaining one.
Setting $a = 0.8$, the black hole metric allows a wide range of parameter
values that yield real and physically meaningful horizons. However, a
comparison with the EHT data shows that only certain values of $\eta$
correspond to the observed shadow sizes, as follows

\begin{itemize}
\item[•] M87$^*$ case: 
\begin{equation*}
0.001 \le \eta \le 0.07, \quad \text{within }1-\sigma,
\end{equation*}
\begin{equation*}
0.001 \le \eta \le 0.09, \quad \text{within }2-\sigma.
\end{equation*}

\item[•]  Sgr~A$^*$ case (EHT$_{\text{VLTI}}$) : 
\begin{equation*}
0.001 \le \eta \le 0.05, \quad \text{within }1-\sigma,
\end{equation*}
\begin{equation*}
0.001 \le \eta \le 0.075, \quad \text{within }2-\sigma.
\end{equation*}

\item[•] Sgr~A$^*$  case (EHT$_{\text{Keck}}$): 
\begin{equation*}
0.001 \le \eta \le 0.06, \quad \text{within }1-\sigma,
\end{equation*}
\begin{equation*}
0.001 \le \eta \le 0.07, \quad \text{within }2-\sigma.
\end{equation*}
\end{itemize}

The results indicate that, when all other parameters are kept fixed, $\eta$
is positive and remains below 0.1 to ensure consistency with the
observations.
 
\begin{figure}[h!]
\centering
\includegraphics[scale=0.45]{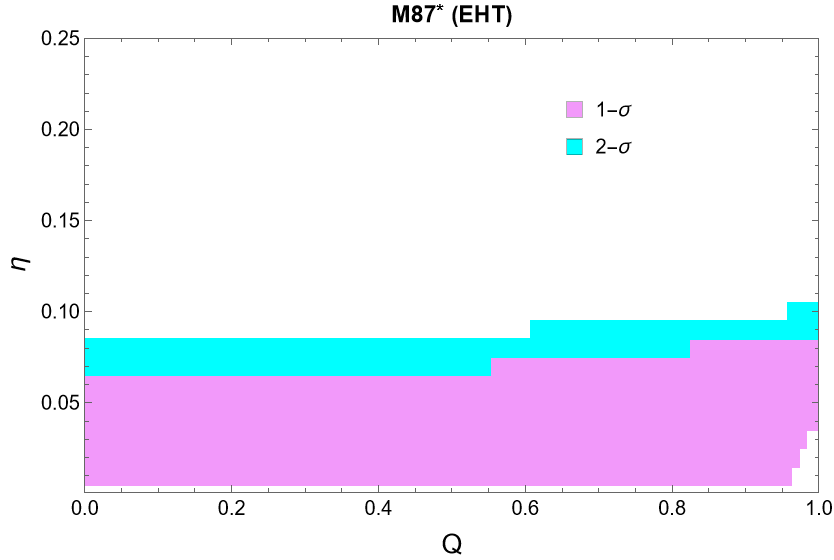}\hspace{2mm} %
\includegraphics[scale=0.45]{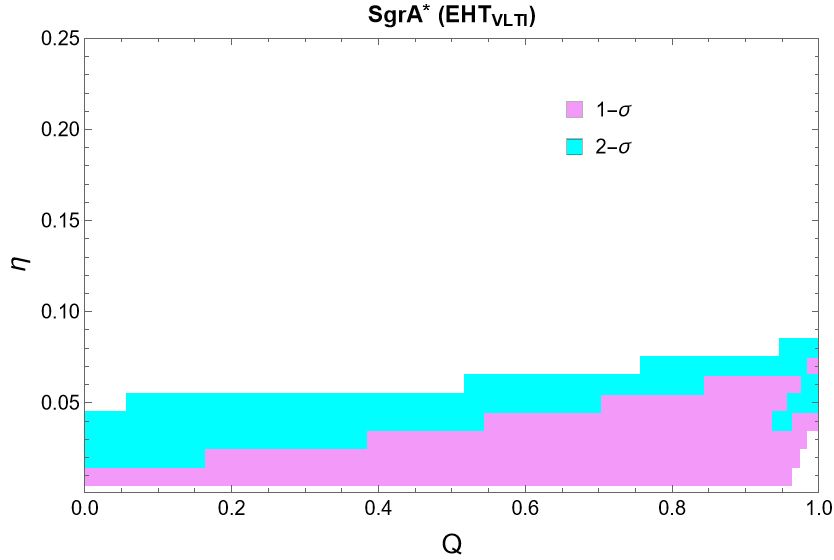} \hspace{2mm} %
\includegraphics[scale=0.45]{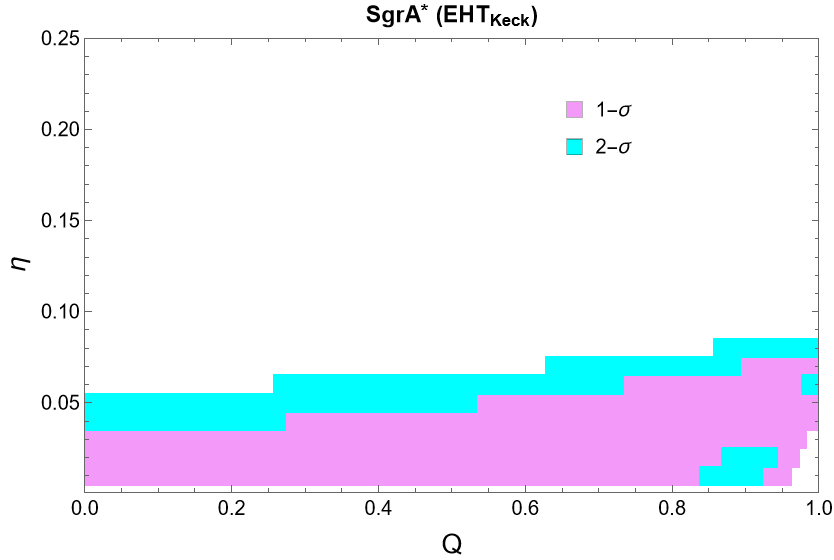}
\caption{ \textit{\protect\footnotesize Constraint Constraint regions in the 
$( \protect\eta,Q)$ plane obtained from CUDA-based simulations, showing
agreement with the EHT observations of M$87^*$ and Sgr~A$^*$ within $1-%
\protect\sigma$ and $2-\protect\sigma$ confidence levels for $a=0.5$ with $%
M=1$.}}
\label{fig:Moduli1}
\end{figure}
Similarly, Fig.~(\ref{fig:Moduli1}) shows the allowed regions in the $%
(\eta,Q)$ plane. Larger values of $\eta$ and $Q$ correspond to a higher
density of points consistent with the observations, suggesting that larger
values of $Q$ improve the agreement between theoretical shadow predictions
and the EHT data. Taking $Q=0.4$, the black hole metric still allows a wide
range of valid horizons. However, only specific values $\eta$ can reproduce
the observed shadows. The corresponding constraints are found to be

\begin{itemize}
\item[•] M87$^*$ case: 
\begin{equation*}
0.001 \le \eta \le 0.065, \quad \text{within } 1-\sigma, 
\end{equation*}
\begin{equation*}
0.001 \le \eta \le 0.085, \quad \text{within } 2-\sigma. 
\end{equation*}

\item[•]  Sgr~A$^*$ case  (EHT$_{\text{VLTI}}$): 
\begin{equation*}
0.001 \le \eta \le 0.035, \quad \text{within } 1-\sigma, 
\end{equation*}
\begin{equation*}
0.001 \le \eta \le 0.055, \quad \text{within } 2-\sigma. 
\end{equation*}

\item[•]  Sgr~A$^*$  case (EHT$_{\text{Keck}}$): 
\begin{equation*}
0.001 \le \eta \le 0.045, \quad \text{within } 1-\sigma, 
\end{equation*}
\begin{equation*}
0.001 \le \eta \le 0.065, \quad \text{within } 2-\sigma. 
\end{equation*}
\end{itemize}

These results confirm that, when all other parameters are held constant, $%
\eta$ should remain positive and less than approximately $0.1$ to be
consistent with the observations.

\section{Conclusion and open questions}

In this paper, we have studied the shadow of rotating charged
Euler--Heisenberg black holes with GMs using high-performance
CUDA numerical codes. First, we have analyzed the horizon structure via the metric function, which
encodes the involved shadow  properties of the solutions. Then, we have
applied the Hamilton--Jacobi formalism together with CUDA-accelerated
simulations to determine the shadow one dimensional curves and the energy emission rates by
varying the black hole parameters. More specifically, we have shown that the
parameter $b$ does not affect either the size of the shadow or the rate of energy emission.  In contrast, we have observed that the rotation, the electric charge,
and the GM parameters. They have influenced both the shadow geometry
and the emission characteristics. Interestingly, for large values of the
rotation parameter and small values of $\eta$, we have observed that the
shadow exhibits a D-like form. However, we have found that this D-like form
gradually disappears as $\eta$ increases, illustrating the interplay between
the GM and the black hole rotation. In fact, the rotation of
such black holes is primarily slowed down by these monopole contributions.

Finally, we have developed a CUDA-based numerical framework to constrain the
black hole parameters by establishing  a direct comparison with astrophysical observations including EHT international collaboration.
Using this framework, we have shown that the GM parameter $\eta$ must
remain positive and below approximately $0.1$   in order to  match  such  observations.

This work raises several questions for future investigations. Specifically,
it would be interesting to study    alternative optical behaviors.  It could be possible to explore the effects of additional
matter fields  including  dark sources or modified interactions on both the shadow and the rate of
energy emission. Such studies could shed further light on the observational
signatures of non-standard gravity models.

\section*{Acknowledgements}

MJ gratefully acknowledges the financial support of the CNRST in the frame
of the PhD Associate Scholarship Program PASS.

\end{document}